\documentclass[12pt,preprint]{aastex}

\shorttitle{Turbulent Cosmic-Ray Accelaration}
\shortauthors{Fujita et al.}

\begin{document}

\title{Turbulent Cosmic-Ray Reacceleration at Radio Relics and Halos in
Clusters of Galaxies}

\author{Yutaka Fujita}
\affil{Department of Earth and Space Science, Graduate School of
 Science, Osaka University, Toyonaka, Osaka 560-0043, Japan}
\email{fujita@vega.ess.sci.osaka-u.ac.jp}

\author{Motokazu Takizawa} 
\affil{Department of Physics, Yamagata
University, 1-4-12 Kojirakawa-machi, Yamagata 990-8560, Japan}

\author{Ryo Yamazaki} 
\affil{Department of Physics and Mathematics,
Aoyama Gakuin University, Fuchinobe, Chuou-ku, Sagamihara 252-5258,
Japan}

\author{Hiroki Akamatsu} 
\affil{SRON Netherlands Institute
for Space Research, Sorbonnelaan 2, 3584 CA Utrecht, The Netherlands}

\and

\author{Hiroshi Ohno} 
\affil{Tohoku Bunkyo College, 515 Katayachi, Yamagata 990-2316, Japan}

\begin{abstract}
 Radio relics are synchrotron emission found on the periphery of galaxy
 clusters. From the position and the morphology, it is often believed
 that the relics are generated by cosmic-ray (CR) electrons accelerated
 at shocks through a diffusive shock acceleration (DSA)
 mechanism. However, some radio relics have harder spectra than the
 prediction of the standard DSA model. One example is observed in the
 cluster 1RXS J0603.3$+$4214, which is often called the ``Toothbrush
 Cluster''. Interestingly, the position of the relic is shifted from
 that of a possible shock. In this study, we show that these
 discrepancies in the spectrum and the position can be solved if
 turbulent (re)acceleration is very effective behind the shock. This
 means that for some relics turbulent reacceleration may be the main
 mechanism to produce high-energy electrons, contrary to the common
 belief that it is the DSA.  Moreover, we show that for efficient
 reacceleration, the effective mean free path of the electrons has to be
 much smaller than their Coulomb mean free path. We also study the
 merging cluster 1E~0657$-$56 or the ``Bullet Cluster'', in which a
 radio relic has not been found at the position of the prominent shock
 ahead of the bullet. We indicate that a possible relic at the shock is
 obscured by the observed large radio halo that is generated by strong
 turbulence behind the shock. We propose a simple explanation if the
 morphological differences of radio emission among the Toothbrush, the
 Bullet, and the Sausage (CIZA~J2242.8$+$5301) Clusters.
\end{abstract}


\section{Introduction}

Radio relics and halos are diffuse sources that are often found in
galaxy clusters. While the former are located on the periphery of a
cluster, the latter are observed at the center of a cluster. They both
seem to be related to an ongoing merger event \citep{fer08a}. Since they
are synchrotron emission, their presence indicates that cosmic-ray (CR)
electrons are widely distributed in clusters. Observationally, the
relics appear to be associated with shocks created by cluster mergers
\citep{gia08a,mac11a,bou13a,ogr13b}, and thus diffusive shock
acceleration (DSA) may be working there
\citep{roe99a,fuj01b,kan12a,yam15a}. On the other hand, the halos may
originate from electrons (re)accelerated by cluster-wide turbulence
excited by cluster mergers \citep{sch87a,bru01a,pet01b,fuj03a}. They may
also be produced by electrons created through hadronic interactions
\citep{min01a,pfr04a,kes10a,ens11a}.

The galaxy cluster 1RXS J0603.3$+$4214 has a famous radio relic called
``toothbrush'' \citep{van12b}. Thus, we call this cluster the
``Toothbrush Cluster'' from now on. The relic is located $\sim 1$~Mpc
away from the cluster center. The Mach number of a possible shock
associated with the relic is less than 2
\citep{ogr13a,ita15a,van15a}. Curiously, the observed radio spectrum is
much harder than the prediction of the standard DSA model based on the
Mach number. Moreover, the position of the relic does not seem to
coincide with that of the shock \citep{ogr13a}. These facts suggest that
the CR electrons responsible for the relic are not accelerated through
the DSA mechanism, although projection effects of multiple shocks may
account for them (\citealt{hon15}; see also \citealt{vaz12,ski13}).

The merging cluster 1E~0657$-$56 is known as the ``Bullet Cluster'',
because a small cluster (``bullet'') is passing through a large cluster
with a velocity of $\gtrsim 4000\rm\: km\:s^{-1}$
\citep{mar02b,mar06c}. X-ray observations show that this cluster has a
prominent shock with a Mach number of ${\cal M}=3$ ahead of the
bullet. Behind the shock, a large radio halo is developing
\citep{lia00a,mar02b,shi14a}. However, despite of the relatively large
Mach number, a radio relic has not been found at the position of the
shock in this cluster. This seems to contradict the fact that ${\cal
M}\sim 3$ is a normal value for the development of a relic at a shock
\citep{aka13a}.

In this paper, we explore the possibility that the relic in the
Toothbrush Cluster is mainly associated with turbulence developed {\it
behind} a shock. This may contradict the common idea that radio relics
are basically generated by the DSA and radio halos are generated by the
turbulent reacceleration.  In our model, CR electrons weakly accelerated
at the shock are significantly reaccelerated in the turbulence
(\citealt{mar05a,mer11a,sas15a}; see also \citealt{ino09a}). In other
words, the DSA at the shock only provides seed CR electrons that are
reaccelerated in the turbulence.  Since the shock is weak, the radio
emission at the shock is not strong. However, if the turbulent
reacceleration is effective enough, the radio emission from the
reaccelerated electrons can be strong, and the spectrum can be harder
than the prediction of the DSA model. Moreover, the difference of the
position of the shock and the relic in the Toothbrush Cluster can
naturally be explained. We also consider the Bullet Cluster as an
extreme case of our model. That is, the reacceleration is so effective
that the emission from the reaccelerated electrons, which is observed as
a radio halo, obscures a possible radio relic at the shock.

\section{Models}

We assume that electrons accelerated at a shock are reaccelerated by
turbulence in the downstream of the shock. For a given upstream gas
density $\rho_{\rm u}$, an upstream velocity $V_{\rm u}$ on the shock
frame, and a Mach number ${\cal M}$, the downstream density and the
velocity are $\rho_{\rm d}=r_c \rho_{\rm u}$ and $V_{\rm d}=V_{\rm
u}/r_c$, respectively, where
\begin{equation}
 r_c = \frac{(\gamma_g+1){\cal M}^2}{(\gamma_g-1){\cal M}^2+2}
\end{equation}
is the shock compression ratio and $\gamma_g=5/3$ is the adiabatic
index. For a given temperature $T_{\rm u}$ of the upstream, that of the
downstream is written as
\begin{equation}
 \frac{T_{\rm d}}{T_{\rm u}} = \frac{[2\gamma_g{\cal M}^2-(\gamma_g-1)]
[(\gamma_g-1){\cal M}^2+2]}{(\gamma_g+1)^2{\cal M}^2}\:.
\end{equation}
We assume a planar shock.

The momentum spectrum of electrons accelerated through the DSA mechanism
at the shock is
\begin{equation}
\label{eq:f0}
 n_0(p)= A_0 p^{-s}\:,
\end{equation}
where
\begin{equation}
 s = \frac{r_c+2}{r_c-1}
\end{equation}
\citep{bla87a}. Here the number density of electrons with momenta
between $p$ and $p+dp$ is $n_0(p)dp$. The lowest momentum of CR
electrons is $p_{\rm min}=m_{\rm e} c$. The normalization $A_0$ is given
as follows. The kinetic energy flux of gas from the upstream of the
shock is $\rho_{\rm u} V_{\rm u}^3/2$. We assume that the fraction
$\chi_{\rm e}$ of that flux becomes the kinetic energy of CR
electrons. Thus, the energy density of CR electrons just downside of the
shock is $\epsilon_{e,sh}=\chi_{\rm e}\rho_{\rm u} V_{\rm u}^3/(2 V_{\rm
d}) = \chi_{\rm e}\rho_{\rm u} V_{\rm u}^2 r_c/2$. On the other hand,
\begin{equation}
 \epsilon_{e,sh}=\int_{p_{\rm min}}^\infty m_{\rm e} c^2 (\gamma-1) n_0(p) dp
\:,
\end{equation}
where $\gamma$ is the Lorentz factor for the particles. Thus,
\begin{equation}
 A_0 = \frac{\chi_{\rm e}\rho_{\rm u} V_{\rm u}^3 r_c}{2 m_{\rm e} c^2}
/\int_{p_{\rm min}}^\infty (\gamma-1) p^{-s} dp \:.
\end{equation}

In the downstream of the shock, electrons are reaccelerated in turbulence
and the spectrum changes as a function of time, $n(t,p)$, where
$n(t=0,p)=n_0(p)$. The evolution is dictated by the Fokker-Planck
equation,
\begin{equation}
\label{eq:evo}
 \frac{\partial n}{\partial t}
-\frac{\partial}{\partial p}\left(p^2 D_{pp}
\frac{\partial}{\partial p}\frac{n}{p^2}\right)
+ \frac{\partial}{\partial p}\left(\frac{dp}{dt}n\right) = 0\:,
\end{equation}
where $D_{pp}$ is the diffusion coefficient in momentum for scattering.
Since CR electrons are swept downstream with thermal gas, the evolution
of the spectrum described by equation~(\ref{eq:evo}) can be interpreted
as the spatial change of the spectrum. The distance from the shock $x$
is represented by the time $t$ as in $x=V_{\rm d} t$. The value of $x$
increases downstream.

We consider resonant acceleration in which Alfv\'en waves scatter
particles because Alfv\'enic turbulence may cascade to a very small
scale (see Section~2.2.2 of \citealt{bru14a}), which reduces an
effective mean free path of particles and increases efficiency of
reacceleration as is shown below. The momentum diffusion coefficient is
\begin{equation}
\label{eq:Dpp}
 D_{pp}\sim \frac{1}{9}p^2 \frac{v_{\rm A}^2}{D_{xx}}\:,
\end{equation}
where $v_{\rm A}$ is the Alfv\'en velocity and $D_{xx}$ is the spatial
diffusion coefficient given by
\begin{equation}
\label{eq:Dxx}
 D_{xx}\sim \frac{c l_{\rm mfp}}{3}\:,
\end{equation}
where $l_{\rm mfp}$ is the electron mean free path (\citealt{ohn02a};
see also \citealt{ise87a,sch89a,fuj03a,bru04a}). In the case of
turbulent acceleration by magnetosonic waves, it is often assumed that
$l_{\rm mfp}$ is the Coulomb mean free path of thermal particles $l_{\rm
mfp,C}$ (e.g. \citealt{zuh13a}; see also \citealt{bru07a}), which
depends on $\rho_{\rm d}$ and $T_{\rm d}$. In this study, we treat $l_{\rm mfp}$ as
a free parameter represented by
\begin{equation}
\label{eq:lmfp}
 l_{\rm mfp}(t,p)=\eta(t) (p/p_0)^{2-q} l_{\rm mfp,C}\:,
\end{equation}
where $\eta(t)$ is the reduction factor ($\eta\leq 1$), $p_0$ is the
reference momentum, and $q$ is the parameter that represents the
property of the turbulence. We fix $p_0$ at $10^4 m_{\rm e} c$ because
electrons around that momentum are responsible for the observed
synchrotron emissions. We assume that $q=5/3$, which represents the
Kolmogorov case.  Equation~(\ref{eq:lmfp}) means that for $p\sim p_0$
and $\eta\ll 1$ the turbulent cascade extends to scales much smaller
than $l_{\rm mfp,C}$ or waves are created by plasma instabilities at the
small scales \citep{bru11a,bru14a}. We consider time (or spatial)
dependence of $\eta$ because the waves responsible for the particle
scattering develop at $x\lesssim L_{\rm t}$ according to the cascade of
the eddies, where $L_{\rm t}$ is the scale comparable to the size of the
largest eddies. Since the evolution of the turbulence should reflect
this scale, $\eta(t)$ evolves on a timescale of $\sim L_{\rm t}/V_{\rm
d}$, which is much larger than the scattering time of a particle $l_{\rm
mfp}/c$. In this study, we do not discuss the microphysics of the
turbulent reacceleration in detail because it is not well known
\citep{sas15a}. Instead, we treat it phenomenologically and obtain the
value and changing rate of $\eta(t)$ that are consistent with
observations of radio profiles.  Moreover, we do not consider the back
reaction of CRs on the turbulence for simplicity because we do not
specify the turbulence on the microscale; the effect is included in
$\eta(t)$ instead. In equation~(\ref{eq:evo}), $dp/dt$ represents
cooling of electrons. We include synchrotron, inverse Compton
scattering, nonthermal bremsstrahlung, and Coulomb interaction. The
synchrotron emission depends on the magnetic field of the downstream of
the shock $B_{\rm d}$.

\section{Results}
\label{sec:result}

\subsection{Toothbrush Cluster}

We assume that $V_{\rm u}=2500\rm\: km\: s^{-1}$, $\rho_{\rm
u}=5.7\times 10^{-28}\rm\: g\: cm^{-3}$, ${\cal M}=1.5$, and $T_{\rm
u}=4$~keV based on the X-ray observations
\citep{ogr13a,ita15a}. Moreover, we assume that $L_{\rm t}=90$~kpc and
$\chi_{\rm e}=2.5\times 10^{-7}$. These are chosen to be consistent with
radio observations (see later). For these values, the Coulomb mean free
path in the downstream is $l_{\rm mfp,C}\sim 19$~kpc. The magnetic field
in the downstream is $B_{\rm d}=2.0\rm\: \mu G$, which is given by
observations \citep{ita15a}.

The functional form of the reduction factor is
\begin{equation}
\label{eq:alpha_t}
 \eta(t) (p/p_0)^{2-q} = \min\{\eta_{\rm min} (p/p_0)^{2-q}
 \exp(t/t_0),1\}\:.
\end{equation}
In the fiducial model for the Toothbrush Cluster, we assume that
$\eta_{\rm min}=1.3\times 10^{-4}$, and $t_0 = L_{\rm t}/V_{\rm
d}$. This change of the mean free path is associated with the decay of
turbulence. We assume that the Alfv\'en velocity behind the shock as
$v_{\rm A}=B_{\rm d}/\sqrt{4\pi\rho_{\rm d}}$. For the parameters we
adopted, the pressure of the magnetic fields just behind of the shock,
$B_{\rm d}^2/(8\pi)$, is smaller than the thermal pressure there. Note
that the parameters $\eta_{\rm min}$ and $B_{\rm d}$ are degenerated for
reacceleration because both smaller $\eta_{\rm min}$ and larger $B_{\rm
d}$ result in higher efficiency of reacceleration
(equations~\ref{eq:Dpp} and~\ref{eq:Dxx}), although $B_{\rm d}$ affects
synchrotron cooling. The parameters $L_{\rm t}$ and $\chi_{\rm e}$
determine the size of the turbulent region and the normalization of the
CR spectrum, respectively. The free parameters in this model are
basically $\eta$ (or $\eta_{\rm min}$ and $L_{\rm t}$) and $\chi_{\rm
e}$.

\begin{figure}
\epsscale{.84} \plottwo{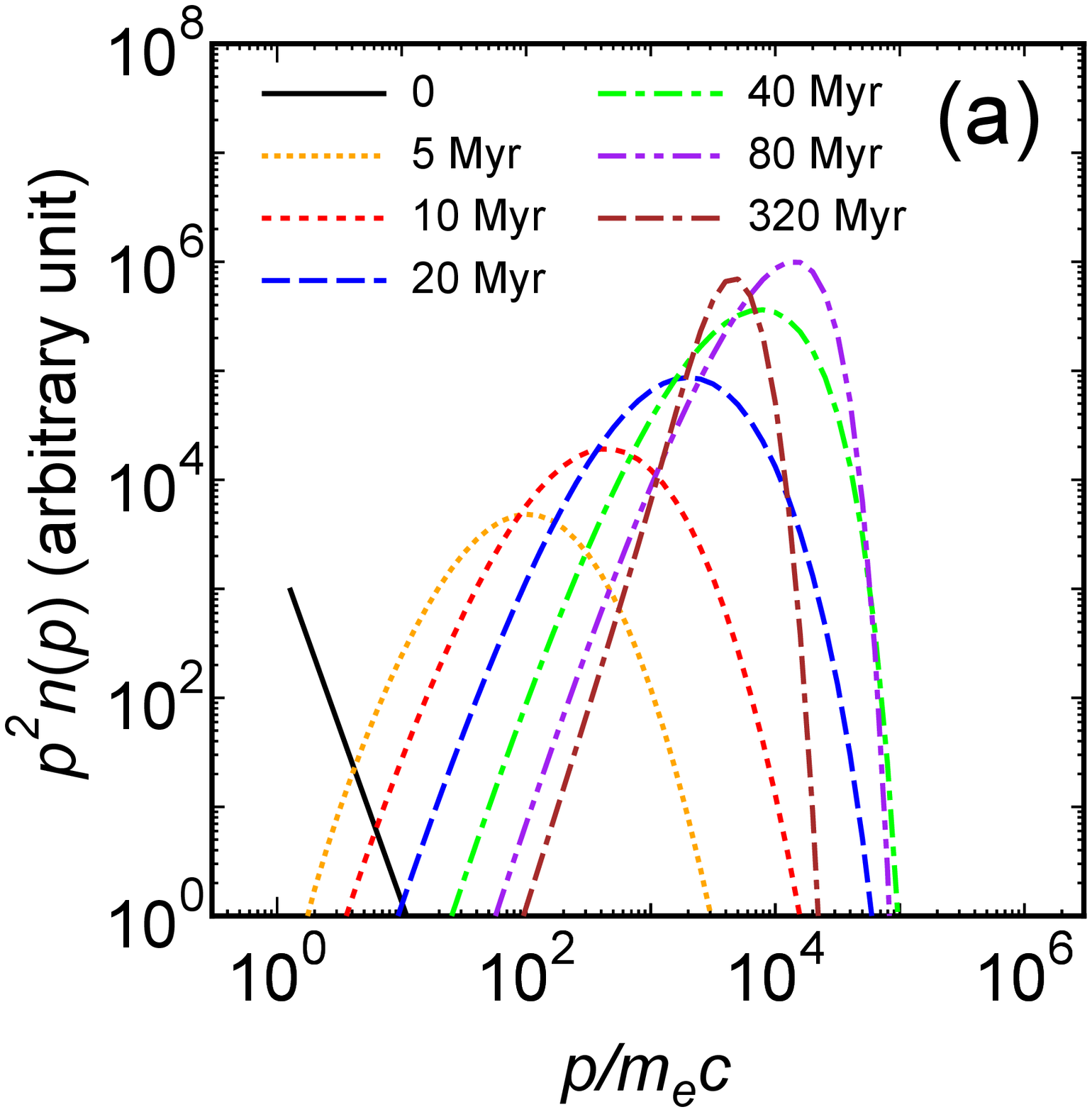}{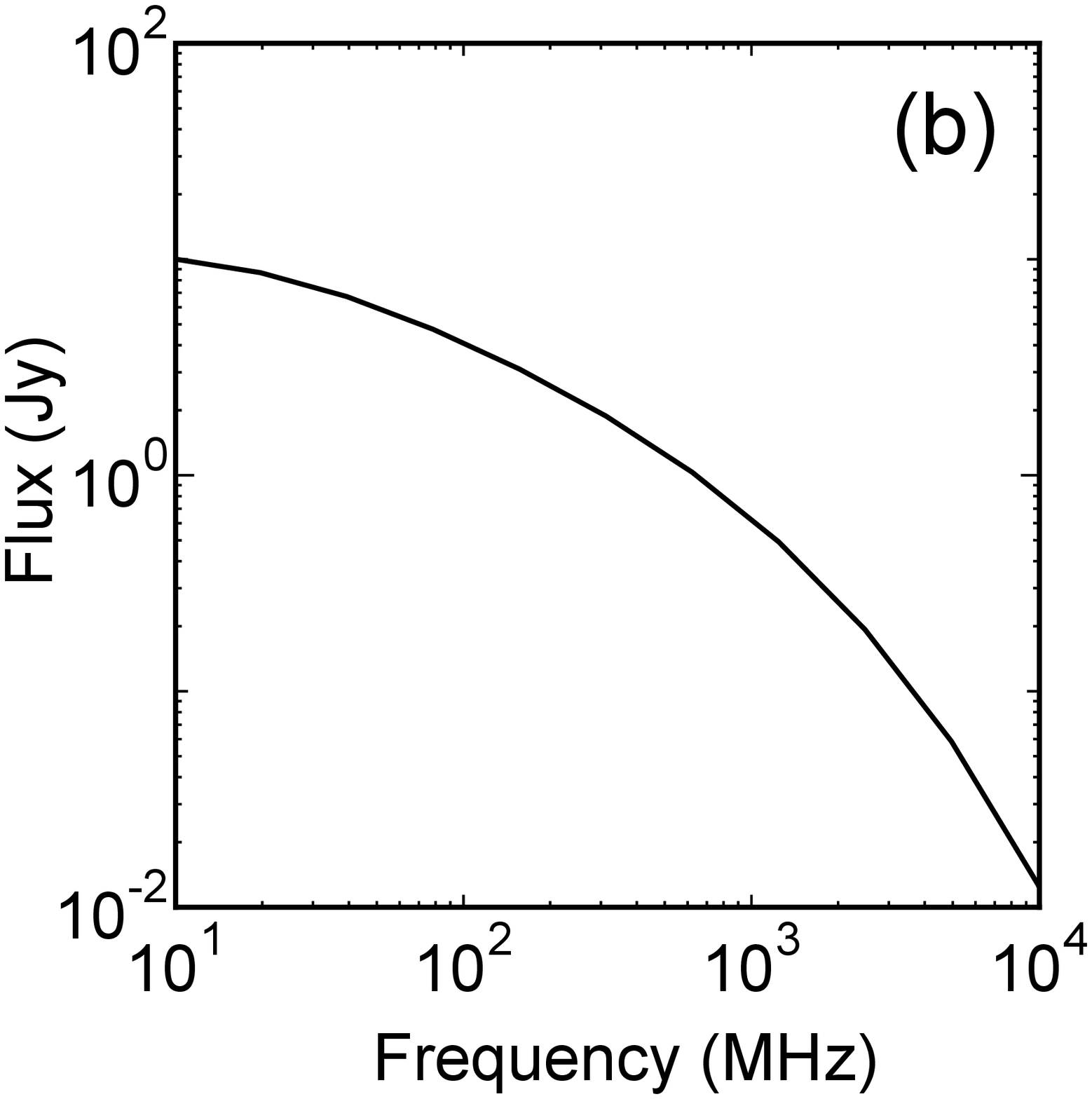} \\ \epsscale{.45}
\plotone{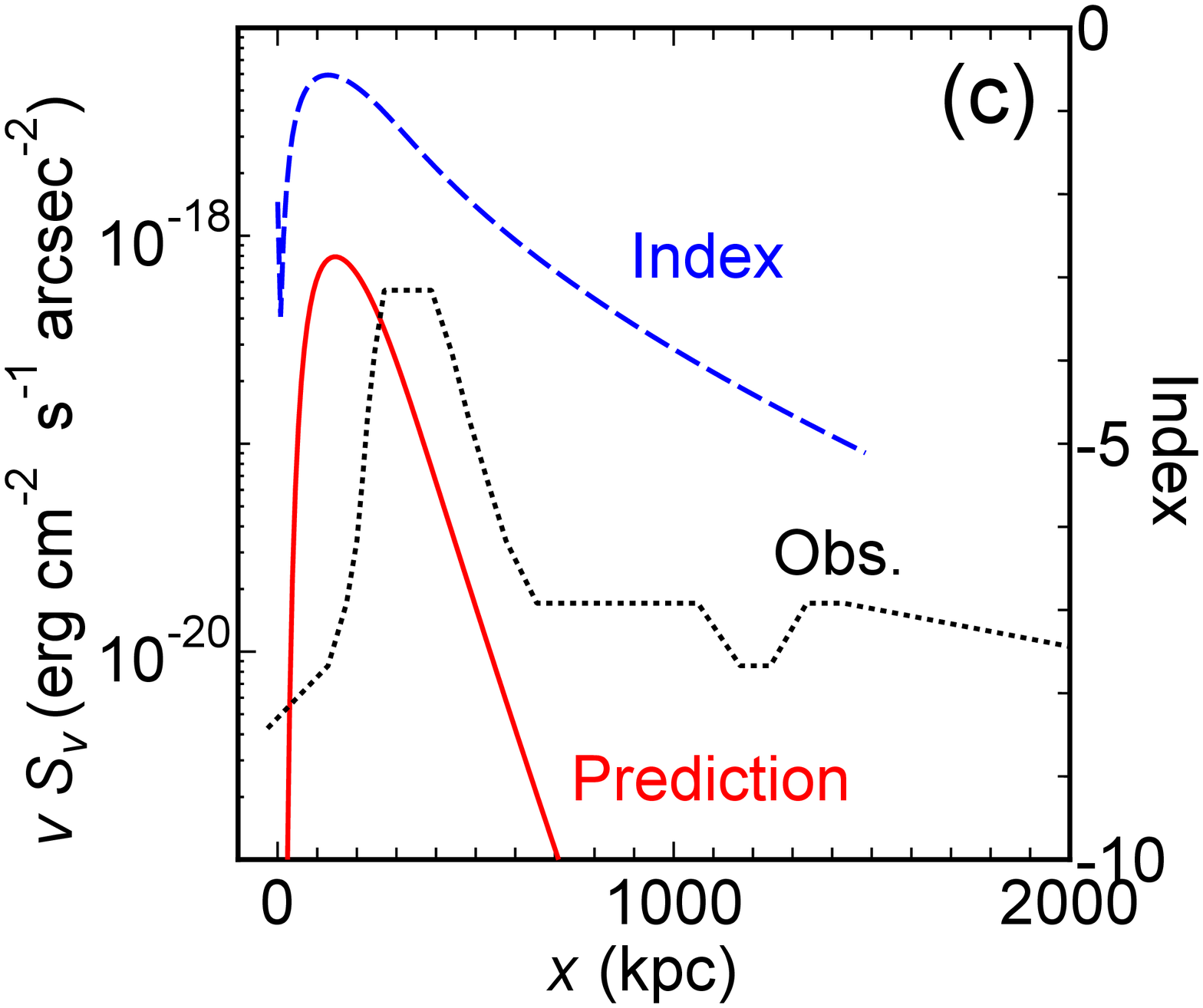} \caption{Results for the Toothbrush Cluster using the
fiducial parameters. (a) Electron spectra behind the shock. Time $t$ is
shown in the legends; $t=320$~Myr corresponds to the distance of
$x=480$~kpc. (b) Integrated radio spectrum. (c) Synchrotron surface
 brightness at 1382~MHz as the function of the distance from the
 shock. The solid curve is our prediction, and the dotted curve is the
 observation \citep{van12b}. The dashed curve is our prediction for the
 spectral index between 150 and 2272~MHz.}\label{fig:tooth}
\end{figure}

Figure~\ref{fig:tooth}a shows the evolution of the electron momentum
spectrum. Initially, the spectrum is soft because of the low Mach number
(${\cal M}=1.5$). In the strong turbulence behind the shock, the
electrons are reaccelerated, which continues until $t\lesssim
t_0=60$~Myr. At this stage, the electron spectrum becomes harder. After
that, the reacceleration weakens owing to the decay of the turbulence
and radiative cooling dominates the reacceleration. As a result, the
number of high-energy electrons ($\gamma\gtrsim 10^4$)
decreases. Figure~\ref{fig:tooth}b presents the integrated synchrotron
flux of the relic. We assume that the relic as a whole is stationary and
both the relic's width and the depth along the line of sight are
500~kpc. The spectrum is not represented by a single power law and is
broadly consistent with the observation \citep{str15a}.

Figure~\ref{fig:tooth}c shows the synchrotron surface brightness at
1382~MHz as the function of the distance from the shock $x$. We assume
that the shock normal is perpendicular to the line of sight and that the
depth of the radio relic along the line of sight is 500~kpc. For
comparison, we show the observation of the west part of the Toothbrush
at the same frequency \citep{van12b}. The position of the shock is set
at the position indicated by the X-ray observations \citep{ogr13a}. Our
model can reproduce the observed sharp increase of the brightness behind
the shock ($x\sim 140$~kpc). The peak of the position is determined by
the acceleration time, $t_{\rm acc}\sim p^2/D_{pp}$. At $x=0$ and
$p=p_0$, the timescale is $t_{\rm acc}=50$~Myr, and the spatial scale is
$x_{\rm acc}(x=0)=V_{\rm d} t_{\rm acc}(x=0)\sim 74$~kpc. Since $t_{\rm
acc}$ increases as $x$ increases and $D_{pp}$ decreases, the position of
the peak ($x\sim 140$~kpc) is a factor of two larger than $x_{\rm
acc}(x=0)$. We note that the position of the peak is closer to the shock
compared with the observations ($x\sim 300$--400~kpc;
\citealt{ogr13a}). This may indicate that it takes some time for the
turbulence to develop behind the shock and that the reacceleration is
delayed. Or there may be an error for the position of the weak shock. In
fact, recent {\it Chandra} observations suggest that the shock is
located just at the northern edge of the relic but not at the peak
\citep{van15a}, which is consistent with Figure~\ref{fig:tooth}c because
the observed radio profile (dotted line) should be shifted to the left
by $\sim 150$--200~kpc. In Figure~\ref{fig:tooth}c, we also plot the
spectral index of the radio emission, which is measured between 147 and
2272~MHz. At $x=0$ ($t=0$), the index is predicted by the standard DSA
model,
\begin{equation}
\label{eq:alpha}
 \alpha_{\rm DSA}=\frac{1}{2}-\frac{{\cal M}^2 + 1}{{\cal M}^2 - 1}
\end{equation}
\citep[e.g.][]{bla87a},and $\alpha_{\rm DSA}=-2.1$ here.  The index
increases up to $\alpha\sim -0.6$ at the peak of the surface brightness,
which is consistent with the observations (Fig.~8 in
\citealt{van12b}). Note that the rapid decrease of the index just behind
the shock is caused by radiative cooling before reacceleration becomes
effective. On the other hand, the diffuse emission observed at $x\gtrsim
700$~kpc cannot be reproduced by our model. This suggests that the
turbulence does not completely decay, and weak reacceleration continues
there. We note that polarization has been detected at the relic of the
Toothbrush Cluster \citep{van12b}. This may be related to compression of
large-scale magnetic fields \citep{iap12a}.

\begin{figure}
\epsscale{.80} \plotone{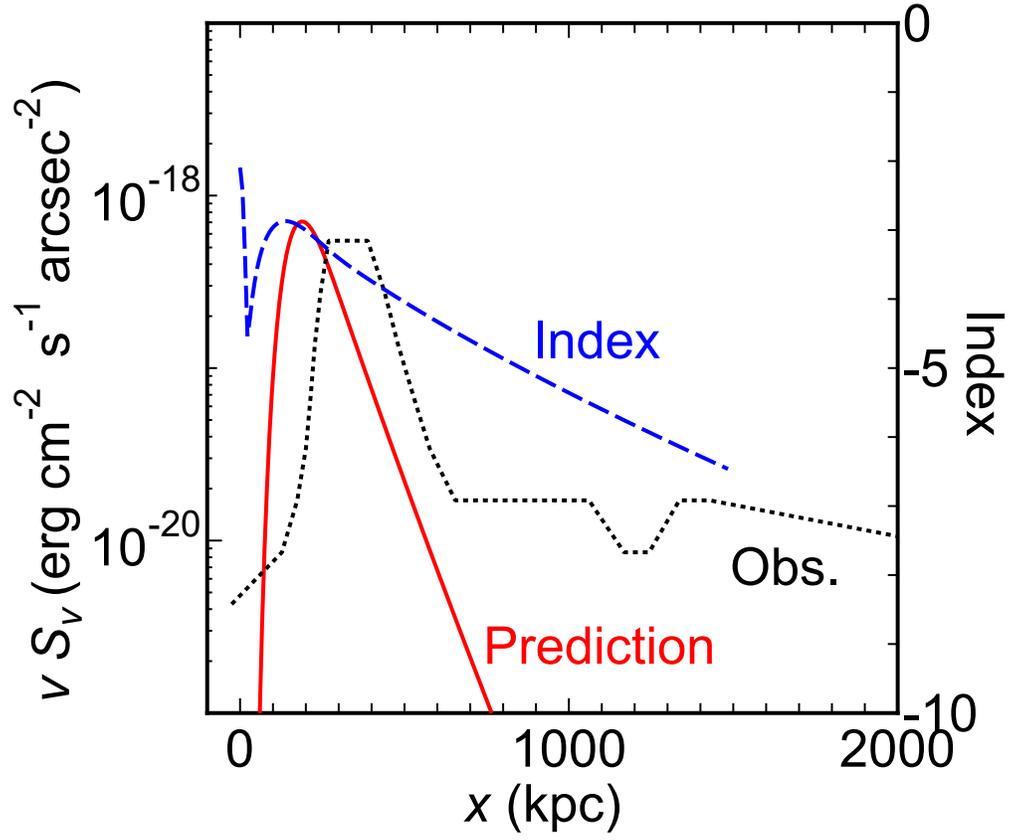} \caption{Same as
Figure~\ref{fig:tooth}(c) but for $\eta_{\rm min}=5.3\times 10^{-4}$ and
$\chi_{\rm e} = 0.01$.}\label{fig:surf_t2}
\end{figure}

Figure~\ref{fig:surf_t2} shows the results when $\eta_{\rm
min}=5.3\times 10^{-4}$ and $\chi_{\rm e}=0.01$; the other parameters
are the same as the fiducial model. The larger $\eta_{\rm min}$ means
that the turbulent reacceleration is less effective than that in the
fiducial model. We compensate the inefficiency with the larger
$\chi_{\rm e}$. Although this model can reproduce the surface brightness
at the peak, the spectral index is much smaller than the observations
($\alpha\sim -0.6$; \citealt{van12b}) because of the weaker
reacceleration. These show that the reduction factor must be small
enough to explain both observations of surface brightness and spectral
index.

\subsection{Bullet Cluster}

The fiducial parameters for the Bullet Cluster are $V_{\rm u}=4700\rm\:
km\: s^{-1}$, $\rho_{\rm u}=1.1\times 10^{-27}\rm\: g\: cm^{-3}$, ${\cal
M}=3.0$, and $T_{\rm u}=9.2$~keV, which are based on the X-ray
observations \citep{mar06c}. Other fiducial parameters are $B_{\rm
d}=30\rm\: \mu G$, and $\chi_{\rm e}=3\times 10^{-10}$. These parameters
are chosen so that the results are consistent with radio observations
(see below).  The magnetic pressure is smaller than the thermal
pressure. Since the radio emission covers the whole cluster
\citep{lia00a,shi14a}, we expect that turbulence is developing in the
whole cluster and the decay scale is much larger than that of the
Toothbrush Cluster. Thus, we simply assume that the reduction factor is
constant ($L_{\rm t}\rightarrow\infty$), and it is $\eta=1\times
10^{-3}$. The Coulomb mean free path in the downstream is $l_{\rm
mfp,C}\sim 190$~kpc.

\begin{figure}
\epsscale{.84} \plottwo{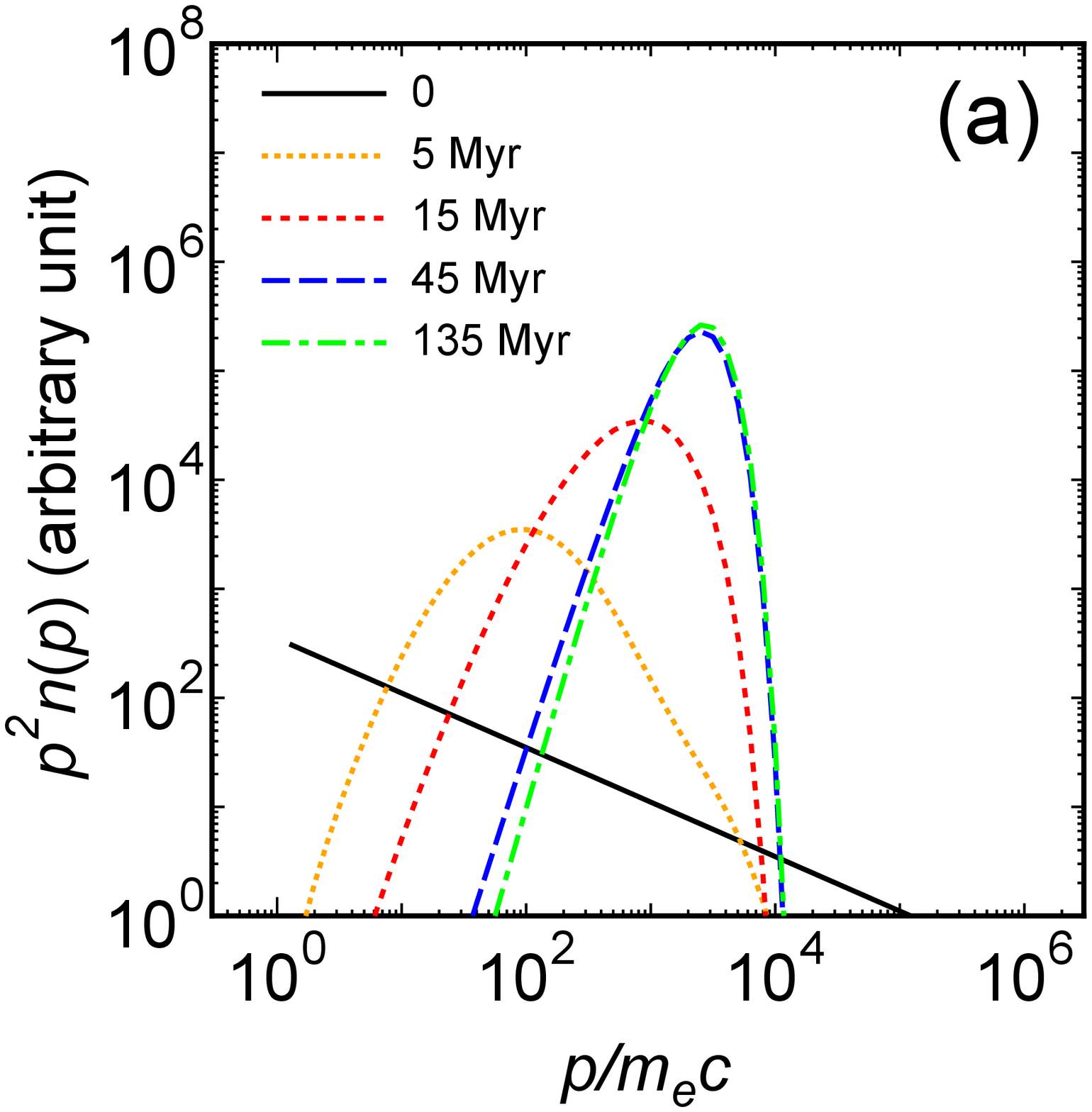}{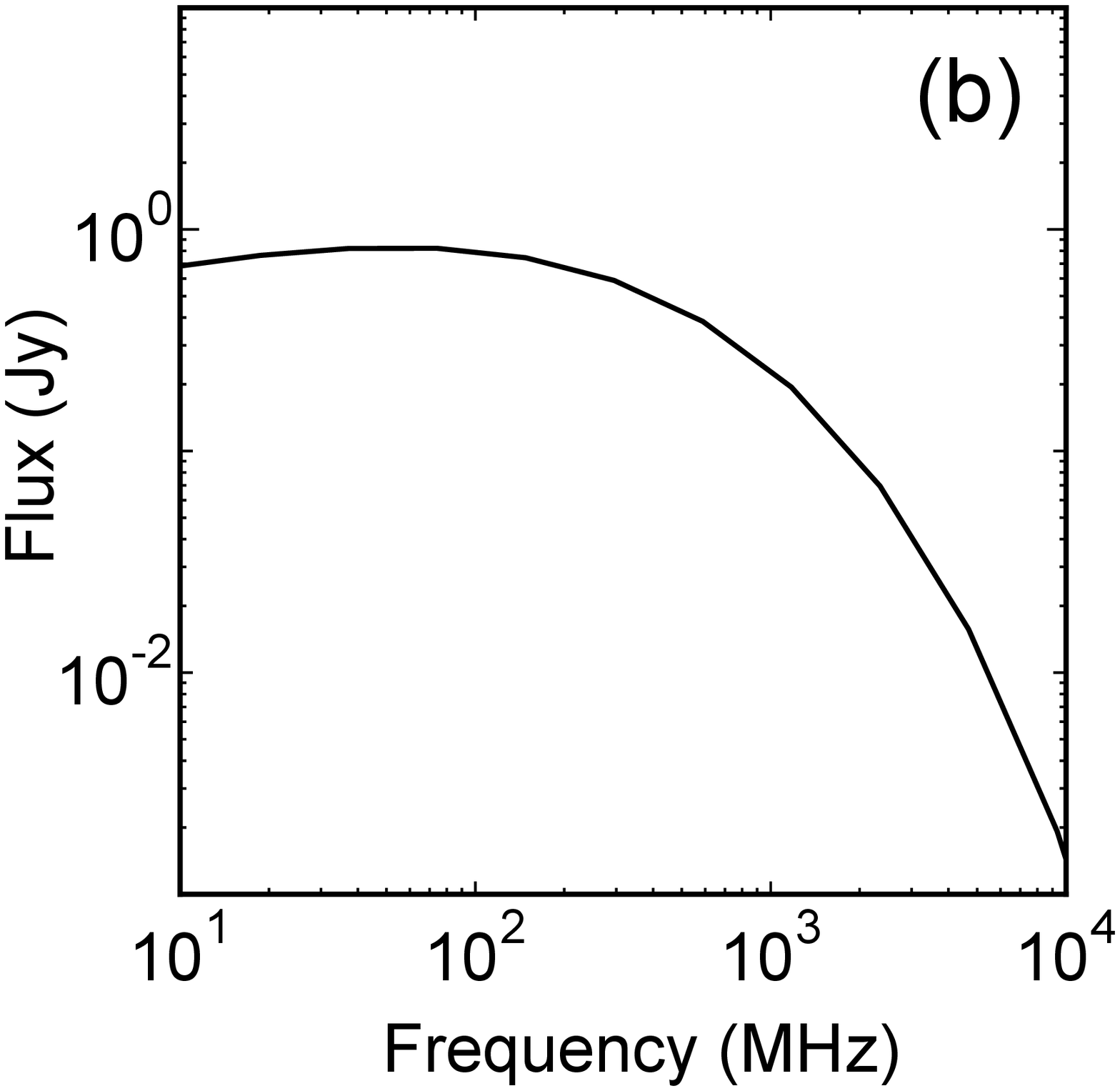} \\ \epsscale{.45}
\plotone{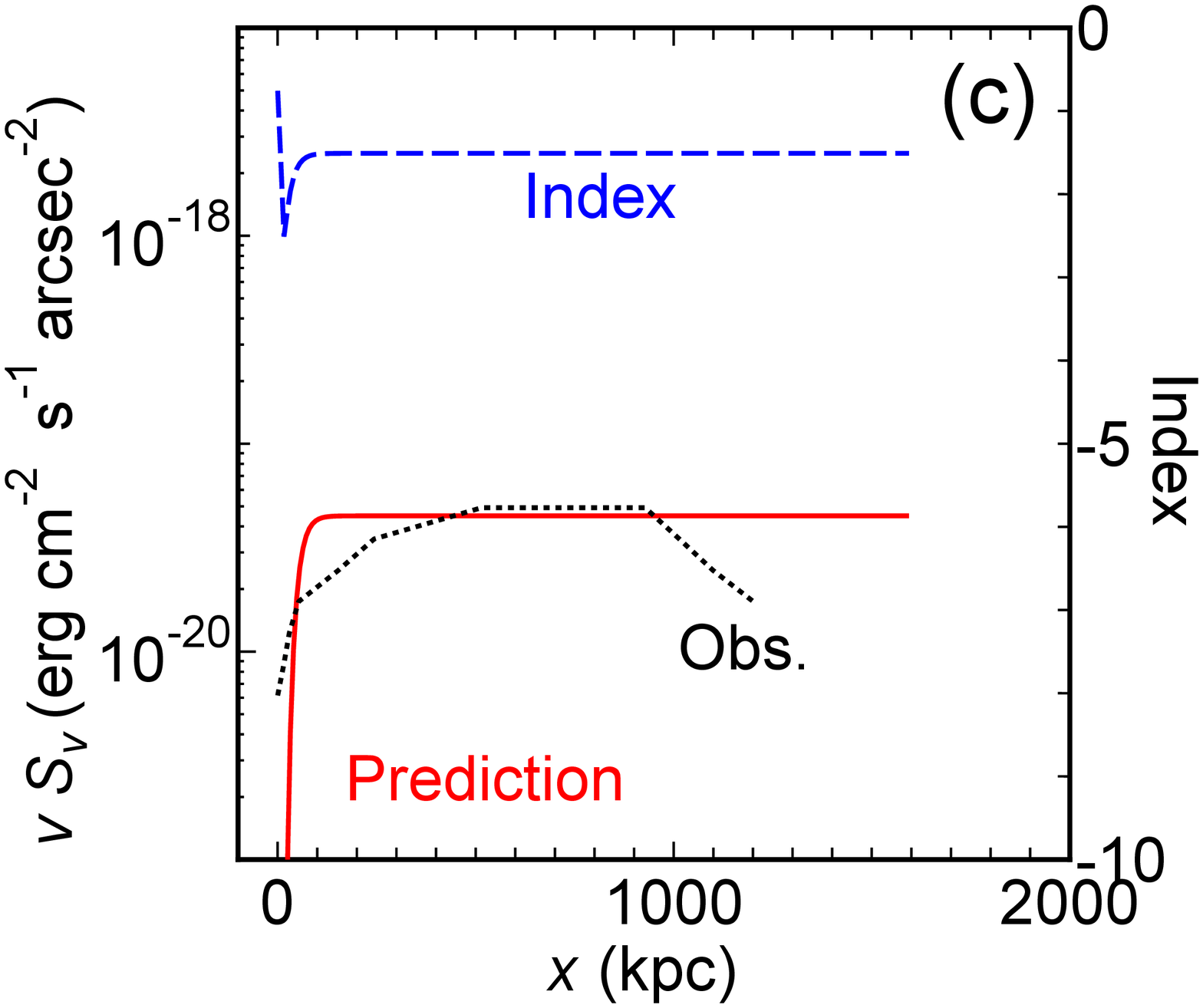} \caption{Results for the Bullet Cluster using the
fiducial parameters. (a) Electron spectra behind the shock. Time $t$ is
shown in the legends; $t=135$~Myr corresponds to the distance of
$x=216$~kpc. (b) Integrated radio spectrum. (c) Synchrotron surface
brightness at 2.1~GHz as a function of the distance from the
shock. The solid curve is our prediction and the dotted curve is the
observation \citep{shi14a}. The dashed curve is our prediction for the
spectral index between 1.1 and 3.1~GHz.}\label{fig:bullet}
\end{figure}

Figure~\ref{fig:bullet}a shows the evolution of the electron momentum
spectrum. Compared to Figure~\ref{fig:tooth}a, the initial spectrum is
harder because of the larger Mach number (${\cal M}=3.0$). As the
electrons are reaccelerated, the peak at $p/(m_{\rm e} c)\sim
10^3$--$10^4$ becomes prominent. Although radiative cooling gradually
becomes effective, turbulent acceleration compensates the effect. For
$t\gtrsim 50$~Myr, the spectrum does not much change because the
acceleration balances with the cooling. Figure~\ref{fig:bullet}b shows
the synchrotron spectrum of the whole cluster, which is not represented
by a power law because it reflects the shape of the momentum spectrum
(Fig.~\ref{fig:bullet}a). We assume that the radio halo's length is
$x=1.6$~Mpc. Both the width and the depth along the line of sight are
500~kpc. The normalization of the synchrotron spectrum depends on the
assumed geometry of the radio-emitting region. Figure~\ref{fig:bullet}c
shows the profile of surface brightness as a function of the distance
from the shock. We simply assume that the position of the shock is at
the tip of the observed radio emission (Fig.~5 of
\citealt{shi14a}). Although our prediction does not perfectly reproduce
the observed profiles, it reproduces the general trend, such as a steep
rise behind the shock and a fairly flat profile after that. We note that
the actual cluster is very complicated. For example, there is a bullet
or a sharp X-ray peak at $x=180$~kpc \citep{shi14a}, and the turbulence
is not expected to be uniform behind the shock. Thus, it would be
useless to fine-tune parameters in our simple model in order to
perfectly reproduce the observations. The spectral index at $x=0$ is
$\alpha_{\rm DSA}=-0.75$ (equation~\ref{eq:alpha}).  For $x\gtrsim
100$~kpc, the index is almost constant (Fig.~\ref{fig:bullet}c). The
average index ($\alpha\sim -1.5$) is consistent with the observations
\citep{shi14a}.

Our results indicate that the surface brightness starts to rise at
$x\sim 30$~kpc (Fig.~\ref{fig:bullet}c), and the gap between the shock
and the rising point is smaller than the resolution of the current radio
telescope ($\sim 100$~kpc; \citealt{shi14a}).  At $x=0$ and $p=p_0$, the
acceleration time and the spacial scale are $t_{\rm acc}=59$~Myr and
$x_{\rm acc}=95$~kpc, respectively, which are comparable to those for
the Toothbrush Cluster.  Observations with a higher resolution in the
future are especially important to reveal the turbulent reacceleration
just behind the shock because the rising point reflects the efficiency
of the reacceleration that depends on parameters such as $\eta$ and
$v_{\rm A} (\propto B_{\rm d})$. The gap is smaller and the spectrum is
harder if $\eta$ is smaller and/or if $v_{\rm A}$ is larger because
acceleration is more efficient.

\begin{figure}
\epsscale{.80} \plotone{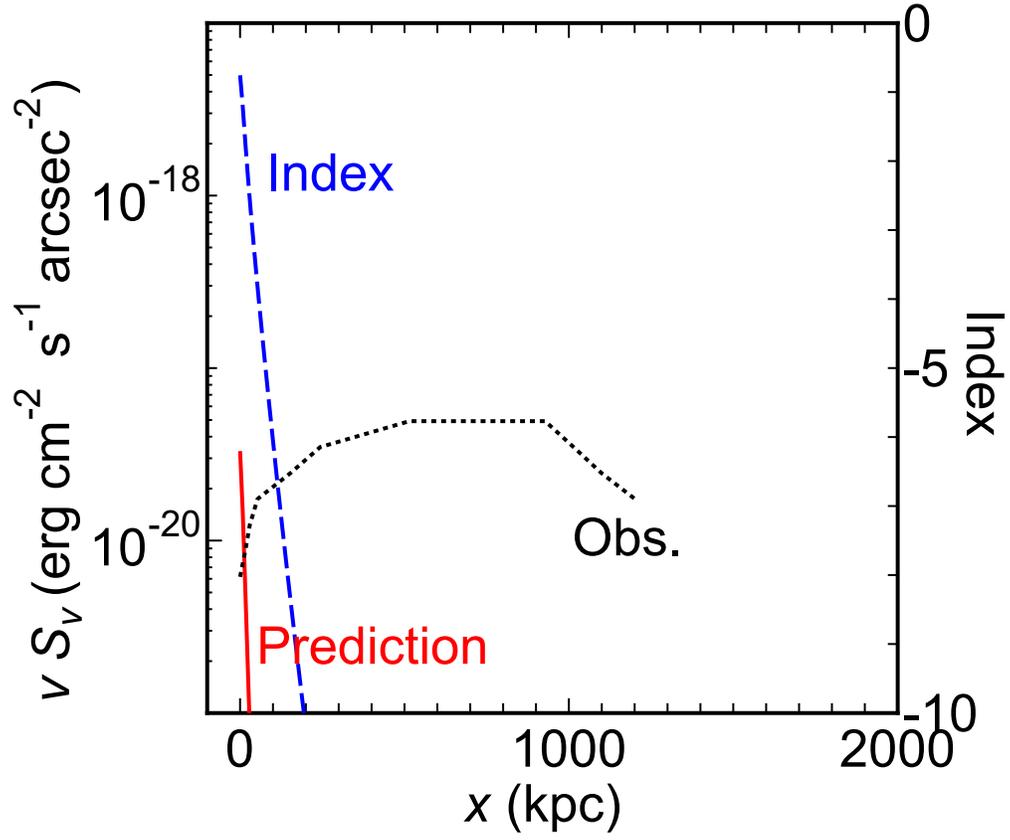} \caption{Same as
Figure~\ref{fig:bullet}(c) but for $\eta=1$ and $\chi_{\rm e} = 3\times 
10^{-6}$.}\label{fig:surf_b2}
\end{figure}

The Mach number of the shock observed in the Bullet Cluster is
relatively large (${\cal M}\sim 3$; \citealt{mar06c}), and it is
comparable to that of the shock observed in CIZA~J2242.8$+$5301
\citep{aka13a,aka15a}, at which a sharp radio relic called ``sausage''
has been observed \citep{van10a}. In contrast with CIZA~J2242.8$+$5301
(the Sausage Cluster hereafter), however, no bow-like radio structure
featuring a relic has been observed in the Bullet Cluster.  This means
that a Mach number is not the only factor that determines radio
morphology. In order to find the reason, we try to reproduce the
sausage-like structure with minimum changes of the parameters from the
fiducial model.  Figure~\ref{fig:surf_b2} shows the profile of the
surface brightness when $\eta=1$ and $\chi_{\rm e}=3\times 10^{-6}$;
other parameters are the same as in the fiducial model for the
Bullet. The larger values of $\eta$ and $\chi_{\rm e}$ mean that while
the turbulent reacceleration is less efficient, the shock acceleration
is more efficient. In Figure~\ref{fig:surf_b2}, the radio emission peaks
at the shock ($x=0$) and rapidly decreases at $x>0$, which are the
properties of the DSA model and are similar to the radio relic in the
Sausage Cluster \citep{van10a}, although the parameters are not chosen
to precisely reproduce the Sausage Cluster. This suggests that the
difference between the Bullet Cluster and the Sausage Cluster reflects
the difference of contribution between the reacceleration in the
turbulence, which depends on $\eta$, and the acceleration at the shock,
which depends on $\chi_{\rm e}$. In other words, the results for the
fiducial parameters (Fig.~\ref{fig:bullet}) indicate that the emission
from the radio halo obscures that from the relic at the shock.

\section{Discussion}

What makes the differences of radio emissions among the Toothbrush
Cluster, the Bullet Cluster, and the Sausage Cluster? The results in
section~\ref{sec:result} show that $\eta$ and $\chi_{\rm e}$ can be the
key parameters to explain them together; the former reflects the
reacceleration efficiency in the turbulence and the latter reflects the
DSA efficiency at the shock.  We speculate that there are three elements
that determine $\eta$ and $\chi_{\rm e}$; development of turbulence, the
Mach number of shocks, and preexisting CRs.

In our study, the difference of radio profiles between the Toothbrush
Cluster (Fig.~\ref{fig:tooth}c) and the Bullet Cluster
(Fig.~\ref{fig:bullet}c) is mainly caused by the difference of
$\eta(t)$, which reflects the decay of turbulence. One possible reason
may be disruption of a cool core. If a fast-moving core is being
violently destroyed through interaction with the ambient medium, strong
turbulence may develop. This may be the case of the Bullet Cluster
\citep{mar06c}. On the other hand, the Toothbrush Cluster does not seem
to have a breaking cool core \citep{ogr13a,van15a}. This may be one
reason that strong turbulence is not developing in the central
region. The mass ratio of subcluster components in a merging cluster may
be a clue. For the Bullet Cluster, the ratio is expected to be around 1
: 0.1 \citep{tak06,aka12}. Weak-lensing analysis showed that the gas
component of the smaller mass component (bullet) is decoupled from its
dark matter component, and the shape of the shock front appears to be
irrelevant to the dark matter distribution \citep{clo04a}. Thus, the
gravitational potential of the smaller component could not hold its gas
and cool core.  The Toothbrush Cluster may be formed through mergers of
three clusters with the mass ratio of 1 : 1 : 0.07 (\citealt{bru12c};
see also \citealt{jee15a}). If their mergers are asymmetric, gas motion
in the cluster may be complicated and possible strong turbulence
associated with the smallest component may not develop at the cluster
center. At the present, only weak turbulence may remain at the center,
and the radio emission from CRs accelerated by the turbulence may be
observed as a diffuse halo at $x\gtrsim 700$~kpc
(Fig.~\ref{fig:tooth}c). For the Sausage Cluster, \citet{oka15a}
recently showed that the mass ratio of two components of the cluster is
1 : 0.5 and that the curved shapes of the two radio relics well match
the density contours of the dark matter. That is, the gas distribution
fairly follows the dark matter distribution. These may indicate that the
mass similarity of the two components may prevent one-sided destruction
of the smaller component and development of strong turbulence. Of
course, if the mass ratio is too large, the smaller component hardly
affects the main cluster. Thus, a moderate mass ratio may be preferable
for development of turbulence.

Our study suggests that the relic of the Toothbrush Cluster is not
basically a result of the DSA, in contrast with the Sausage
Cluster. Their different Mach numbers might be responsible for that. For
example, \citet{ryu03a} indicated that CR acceleration efficiency drops
at ${\cal M}\lesssim 3$. Thus, the relic of the Toothbrush Cluster
(${\cal M}\sim 1.5$) may not satisfy a necessary condition of effective
DSA acceleration (${\cal M}\gtrsim 3$). On the other hand, the Mach
numbers for the Bullet Cluster and the Sausage Cluster satisfy this
condition. Their difference of radio morphology may reflect existence or
nonexistence of seed CR electrons {\it ahead} of the shocks
\citep{ens98a,kan12a,pin13a}. The normalization of the CR spectrum at
the shock, $\chi_{\rm e}$, for Figure~\ref{fig:surf_b2} is $10^4$ times
larger than that for Figure~\ref{fig:bullet}. This may mean that
effective DSA acceleration is required for a relic to be bright at the
shock like the Sausage Cluster (Fig.~\ref{fig:surf_b2}). The
preexisting CR electrons may enhance the efficiency
\citep{kan12a,pin13a}. On the other hand, if there are few preexisting
CR electrons, the efficiency is small and the radio emission is dim at
the shock. However, if strong turbulence is developed behind the shock,
even the electrons with small energies are effectively reaccelerated and
may form a radio halo (Fig.~\ref{fig:bullet}).
 
\section{Summary}

We have investigated the origin of radio relics on the periphery of
galaxy clusters. We focus on the turbulent reacceleration of CR
electrons that had been weakly accelerated at a shock. We considered
Alfv\'en waves as scatterers of the CR electrons. We found that the
effective mean free path of the particles must be much smaller than the
Coulomb mean free path for efficient reacceleration. This may be
realized if the turbulence cascades to the smaller scale or plasma
instabilities excite waves at that small scale. Using our model, we
reproduced the hard spectrum of the relic observed in the cluster 1RXS
J0603.3$+$4214 (the Toothbrush Cluster), which cannot be explained by
the standard DSA model. In our model, the position of the relic does not
coincide with that of the shock. The difference of the positions has
actually been observed in the Toothbrush Cluster. Our results show that
for some relics turbulent reacceleration may be the main mechanism to
produce high-energy electrons, although the turbulent reacceleration is
generally associated with giant radio halos. We also applied our model
to the merging cluster 1E~0657$-$56 (the Bullet Cluster), in which a
radio relic has not been found in spite of the large Mach number of the
bow shock. We showed that strong turbulence behind the shock can
reaccelerate CR electrons. The bright synchrotron emission from those
electrons is observed as a radio halo and it obscures a possible radio
relic at the shock. If the acceleration of the shock were more efficient
and the reacceleration in the turbulence were less efficient, a sharp
radio relic should have been observed in the Bullet Cluster as 
in CIZA 2242.8$+$5301 (the Sausage Cluster). Our results suggest that
various radio relics and halos in clusters can be explained all together
by only two factors: the reacceleration efficiency in the turbulence and
the DSA efficiency at the shock.  We speculate that development of
turbulence, the Mach number of shocks, and preexisting CRs determine
those two factors.

\acknowledgments

We are grateful to R.~J. van Weeren and S.~S. Kimura for useful
comments. This work was supported by KAKENHI No.~15K05080 (Y.F.),
26400218 (M.T.), and 15K05088 (R.Y.). H.A. is supported by a
Grant-in-Aid for Japan Society for the Promotion of Science (JSPS)
Fellows (26-606).

\clearpage

\begin{thebibliography}{}

\bibitem[Akahori 
\& Yoshikawa(2012)]{aka12} Akahori, T., \& Yoshikawa, K.\ 2012, \pasj,
		64, 12

\bibitem[Akamatsu 
\& Kawahara(2013)]{aka13a} Akamatsu, H., \& Kawahara, H.\ 2013, \pasj,
		65, 16

\bibitem[Akamatsu et 
al.(2015)]{2015A&A...582A..87A} Akamatsu, H., van Weeren, R.~J., Ogrean,
		G.~A., et al.\ 2015, \aap, 582, A87

\bibitem[Blandford 
\& Eichler(1987)]{bla87a} Blandford, R., \& Eichler, D.\
		1987, \physrep, 154, 1 

\bibitem[Bourdin et al.(2013)]{bou13a} Bourdin, H., Mazzotta, 
P., Markevitch, M., Giacintucci, S., \& Brunetti, G.\ 2013, \apj, 764,
		82

\bibitem[Br{\"u}ggen et al.(2012)]{bru12c} Br{\"u}ggen, M.,
van Weeren, R.~J., R\"ottgering, H.~J.~A.\ 2012, \mnras, 425, L76 

\bibitem[Brunetti et al.(2004)]{bru04a} Brunetti, G., Blasi, 
P., Cassano, R., \& Gabici, S.\ 2004, \mnras, 350, 1174 

\bibitem[Brunetti 
\& Jones(2014)]{bru14a} Brunetti, G., \& Jones, T.~W.\ 2014,
		International Journal of Modern Physics D, 23, 1430007


\bibitem[Brunetti 
\& Lazarian(2007)]{bru07a} Brunetti, G., \& Lazarian, A.\ 2007, \mnras,
		378, 245

\bibitem[Brunetti 
\& Lazarian(2011)]{bru11a} Brunetti, G., \& Lazarian, A.\
		2011, \mnras, 412, 817 

\bibitem[Brunetti et al.(2001)]{bru01a} Brunetti, G., Setti, 
G., Feretti, L., \& Giovannini, G.\ 2001, \mnras, 320, 365 

\bibitem[Clowe et al.(2004)]{clo04a} Clowe, D., Gonzalez, A., 
\& Markevitch, M.\ 2004, \apj, 604, 596 

\bibitem[Ensslin et 
al.(1998)]{ens98a} Ensslin, T.~A., Biermann, P.~L., Klein, U., \& Kohle,
		S.\ 1998, \aap, 332, 395

\bibitem[En{\ss}lin et 
al.(2011)]{ens11a} En{\ss}lin, T., Pfrommer, C., Miniati, F., \&
		Subramanian, K.\ 2011, \aap, 527, A99

\bibitem[Ferrari et al.(2008)]{fer08a} Ferrari, C., Govoni, 
F., Schindler, S., Bykov, A.~M., \& Rephaeli, Y.\ 2008, \ssr, 134, 93 

\bibitem[Fujita 
\& Sarazin(2001)]{fuj01b} Fujita, Y., \& Sarazin, C.~L.\ 2001, \apj,
		563, 660

\bibitem[Fujita et al.(2003)]{fuj03a} Fujita, Y., Takizawa, 
M., \& Sarazin, C.~L.\ 2003, \apj, 584, 190 

\bibitem[Giacintucci et 
al.(2008)]{gia08a} Giacintucci, S., Venturi, T., Macario, G., et al.\
		2008, \aap, 486, 347

\bibitem[Hong et al.(2015)]{2015ApJ...812...49H} Hong, S.~E., Kang, H., 
\& Ryu, D.\ 2015, \apj, 812, 49 

\bibitem[Inoue et al.(2009)]{ino09a} Inoue, T., Yamazaki, R., 
\& Inutsuka, S.\ 2009, \apj, 695, 825 

\bibitem[Iapichino 
\& Br{\"u}ggen(2012)]{iap12a} Iapichino, L., \& Br{\"u}ggen, M.\ 2012,
		\mnras, 423, 2781

\bibitem[Isenberg(1987)]{ise87a} Isenberg, P.~A.\ 1987, \jgr, 
92, 1067 

\bibitem[Itahana et al.(2015)]{ita15a} Itahana, M., Takizawa, 
M., Akamatsu, H., et al.\ 2015, \pasj~in press, arXiv:1508.05845

\bibitem[Jee et al.(2015)]{jee15a} Jee, M.~J., Dawson, W.~A., 
Stroe, A., et al.\ 2015, arXiv:1510.03486 

\bibitem[Kang et al.(2012)]{kan12a} Kang, H., Ryu, D., 
\& Jones, T.~W.\ 2012, \apj, 756, 97 

\bibitem[Keshet 
\& Loeb(2010)]{kes10a} Keshet, U., \& Loeb, A.\ 2010, \apj, 722, 737 

\bibitem[Liang et al.(2000)]{lia00a} Liang, H., Hunstead, 
R.~W., Birkinshaw, M., \& Andreani, P.\ 2000, \apj, 544, 686 

\bibitem[Macario et al.(2011)]{mac11a} Macario, G., 
Markevitch, M., Giacintucci, S., et al.\ 2011, \apj, 728, 82 

\bibitem[Markevitch(2006)]{mar06c} Markevitch, M.\ 2006, The 
X-ray Universe 2005, 604, 723, arXiv:astro-ph/0511345

\bibitem[Markevitch et al.(2002)]{mar02b} Markevitch, M., 
Gonzalez, A.~H., David, L., et al.\ 2002, \apjl, 567, L27 

\bibitem[Markevitch et al.(2005)]{mar05a} Markevitch, M., 
Govoni, F., Brunetti, G., \& Jerius, D.\ 2005, \apj, 627, 733 

\bibitem[Mertsch \& Sarkar(2011)]{mer11a} Mertsch, P., \&
Sarkar, S.\ 2011, Physical Review Letters, 107, 091101

\bibitem[Miniati et al.(2001)]{min01a} Miniati, F., Jones, 
T.~W., Kang, H., \& Ryu, D.\ 2001, \apj, 562, 233 

\bibitem[Ogrean 
\& Br{\"u}ggen(2013)]{ogr13b} Ogrean, G.~A., \& Br{\"u}ggen, M.\ 2013,
		\mnras, 433, 1701

\bibitem[Ogrean et al.(2013)]{ogr13a} Ogrean, G.~A., 
Br{\"u}ggen, M., van Weeren, R.~J., et al.\ 2013, \mnras, 433, 812 

\bibitem[Ohno et al.(2002)]{ohn02a} Ohno, H., Takizawa, M., 
\& Shibata, S.\ 2002, \apj, 577, 658 

\bibitem[Okabe et al.(2015)]{oka15a} Okabe, N., Akamatsu, H., 
Kakuwa, J., et al.\ 2015, \pasj~in press, arXiv:1508.04558

\bibitem[Petrosian(2001)]{pet01b} Petrosian, V.\ 2001, \apj, 
557, 560 

\bibitem[Pfrommer 
\& En{\ss}lin(2004)]{pfr04a} Pfrommer, C., \& En{\ss}lin, T.~A.\ 2004,
		\mnras, 352, 76

\bibitem[Pinzke et al.(2013)]{pin13a} Pinzke, A., Oh, S.~P., 
\& Pfrommer, C.\ 2013, \mnras, 435, 1061 

\bibitem[Roettiger et al.(1999)]{roe99a} Roettiger, K., Burns, 
J.~O., \& Stone, J.~M.\ 1999, \apj, 518, 603 

\bibitem[Ryu et al.(2003)]{ryu03a} Ryu, D., Kang, H., Hallman, 
E., \& Jones, T.~W.\ 2003, \apj, 593, 599 

\bibitem[Sasaki et al.(2015)]{2015ApJ...814...93S} Sasaki, K., Asano,
K., \& Terasawa, T.\ 2015, \apj, 814, 93

\bibitem[Schlickeiser(1989)]{sch89a} Schlickeiser, R.\ 1989, 
\apj, 336, 243 

\bibitem[Schlickeiser et 
al.(1987)]{sch87a} Schlickeiser, R., Sievers, A., \& Thiemann, H.\ 1987,
		\aap, 182, 21

\bibitem[Shimwell et al.(2014)]{shi14a} Shimwell, T.~W., 
Brown, S., Feain, I.~J., et al.\ 2014, \mnras, 440, 2901 


\bibitem[Skillman et al.(2013)]{ski13} Skillman, S.~W., Xu, 
H., Hallman, E.~J., et al.\ 2013, \apj, 765, 21 

\bibitem[Stroe et al.(2016)]{2016MNRAS.455.2402S} Stroe, A., Shimwell,
T., Rumsey, C., et al.\ 2016, \mnras, 455, 2402

\bibitem[Takizawa(2006)]{tak06} Takizawa, M.\ 2006, \pasj, 
58, 925 

\bibitem[van Weeren et 
al.(2015)]{van15a} van Weeren, R.~J., et al.\ 2015, submitted to ApJ

\bibitem[van Weeren et al.(2010)]{van10a} van Weeren, R.~J., 
R{\"o}ttgering, H.~J.~A., Br{\"u}ggen, M., 
\& Hoeft, M.\ 2010, Science, 330, 347 

\bibitem[van Weeren et 
al.(2012)]{van12b} van Weeren, R.~J., R{\"o}ttgering,
	  H.~J.~A., Intema, H.~T., et al.\ 2012, \aap, 546, A124 


\bibitem[Vazza et al.(2012)]{vaz12} Vazza, F., Br{\"u}ggen, 
M., van Weeren, R., et al.\ 2012, \mnras, 421, 1868 

\bibitem[Yamazaki 
\& Loeb(2015)]{yam15a} Yamazaki, R., \& Loeb, A.\ 2015, \mnras, 453,
		1990

\bibitem[ZuHone et al.(2013)]{zuh13a} ZuHone, J.~A., 
Markevitch, M., Brunetti, G., \& Giacintucci, S.\ 2013, \apj, 762, 78 


\end{thebibliography}
\end{document}